\documentclass[12pt,a4paper]{article}
\usepackage[a4paper]{geometry}
\usepackage[utf8]{inputenc} 
\usepackage[english]{babel}
\usepackage{epsfig}
\usepackage{amsmath, amsfonts, amstext, amscd, bezier,dsfont,bm}
\usepackage{amssymb,amsthm}
\usepackage{indentfirst}
\usepackage{graphicx}
\usepackage{color}
\usepackage{hyperref}
\usepackage[affil-it]{authblk}
\usepackage{multirow}
\usepackage{algorithm}
\usepackage{algorithmicx}
\usepackage{algpseudocode}
\usepackage{appendix}

\newcommand{\A}{\mathbf{A}}
\newcommand{\W}{\mathbf{W}}
\newcommand{\Z}{\mathbf{Z}}
\newcommand{\npi}{\boldsymbol{\pi}}
\newcommand{\nt}{\boldsymbol{\tau}}

\newcommand{\nLambda}{\boldsymbol{\Lambda}}

\newcommand{\ntheta }{\boldsymbol{\theta}}
\newcommand{\nmu}{\boldsymbol{\mu}}
\newcommand{\nnu }{\boldsymbol{\nu}}
\newcommand{\p }{\boldsymbol{p}}
\newcommand{\E }{\mathbb{E}}

\renewcommand{\P }{\mathbb{P}}
\newcommand{\KL }{D_{\mbox{\tiny{KL}}}}
\usepackage{caption,subcaption}
\usepackage{float}

\usepackage{amssymb} 
\usepackage{natbib}

\usepackage{mathtools}
\usepackage{mathrsfs} 
\usepackage{seqsplit}
\usepackage{enumitem}

\theoremstyle{plain}

\theoremstyle{definition}

\theoremstyle{remark}

\numberwithin{equation}{section}
\numberwithin{theorem}{section}

\providecommand{\keywords}[1]{\textit{Keywords:} #1}

\usepackage{mathtools}
\mathtoolsset{showonlyrefs} 

\begin{document}
\title{\Large{Modeling sparsity in count-weighted networks}}

\author[1]{Andressa Cerqueira\thanks{Corresponding author: acerqueira@ufscar.br}}
\author[1]{Laila L. S. Costa}

\affil[1]{Department of Statistics, Universidade Federal de S\~ao Carlos, Brazil}

\date{\today}
\maketitle

\begin{abstract}
Community detection methods have been extensively studied to recover communities structures in network data. While many models and methods focus on binary data, real-world networks also present the strength of connections, which could be considered in the network analysis. We propose a probabilistic model for generating weighted networks that allows us to control network sparsity and incorporates degree corrections for each node. We propose a community detection method based on the Variational Expectation-Maximization (VEM) algorithm. We show that the proposed method works well in practice for simulated networks. We analyze the Brazilian airport network to compare the community structures before and during the COVID-19 pandemic.
\end{abstract}

\keywords{
weighted networks, variational inference, zero inflated distribution
}

\section{Introduction}

Community detection is one of the most studied problems in the analysis of network data. The Stochastic Block Model (SBM), proposed by \cite{holland1983stochastic}, provides a probabilistic framework for modeling community structures in networks by defining the probability of edge existence between nodes based on their community memberships. Since this seminal work, several extensions of the SBM have been proposed, such as, degree corrected SBM \citep{karrer2011}, mixed membership \citep{airoldi08mixed}, overlaping SBM \citep{Latouche2011overlap}, dynamic SBM \citep{Xu2014dynamic,matias2027} and SBM with nodes attributes \citep{Zhang2016attributes}.

Many of these models are proposed for binary networks, where the only available information is whether an edge exists between nodes. However, in many real-world networks, the strength of the connection between two vertices is also important. For example, in airport networks, where nodes represent airports and edges represent flight routes, information about the volume of passengers or flights can also be incorporated into the network structure. Following events such as the COVID-19 pandemic, the routes may remain unchanged, but the traffic between airports can shift dramatically. In such cases, while the binary network structure stays constant, the weighted network can evolve over time. 

In order to model networks with community structure and discrete valued edges, \cite{Mariadassou2010}  proposed a weighted SBM based on a Poisson distribution. However, incorporating network sparsity is crucial when modeling real-world networks, as many of them exhibit a large proportion of zero-valued edges. In fact, in a weighted network, an observed zero can be interpreted in two different ways: either as the absence of an edge or as the existence of an edge with a weight of zero.

In the context of weighted SBMs, \cite{aicher2014} proposed a model that addresses network sparsity by introducing a Bernoulli variable to model edge existence and using an exponential family distribution to model edge strength. This model assumes that the parameter controlling network sparsity does not depend on the community membership of the nodes. To incorporate heterogeneity in the network, \cite{motalebi2021hurdle} proposed a hurdle SBM model, where edge existence is modeled by a degenerate distribution, and edge weights follow a truncated Poisson distribution. In the study of multilayer networks, \cite{dong2020modeling} proposed the multivariate zero-inflated Poisson SBM to model correlations among different layers. The single-layer special case of this model can describe discrete-valued weighted networks; however, it does not account for node degree heterogeneity.

To take into account these characteristics of the network, we propose a variation of the SBM that models sparsity in discrete edge-weighted networks. Specifically, we use the zero-inflated Poisson (ZIP) distribution to model edge strength, combining two processes for generating zero values: one for structural zeros and another for the variability of positive edge weights. Since the classical SBM tends to create communities of nodes with similar degrees, incorporating degree correction is necessary for accurately modeling graphs with heterogeneous degree distributions. Therefore, we incorporate degree corrections into this model as proposed by \cite{karrer2011}, and extended to directed dynamic networks by \cite{riverain2023poisson}.

The community detection method employed in this work is based on likelihood maximization. When using likelihood methods for estimation in latent models, the Expectation-Maximization (EM) algorithm is typically employed. However, it is well known that the EM algorithm becomes intractable for certain variants of the SBM model \cite{daudin2008mixture}. As an alternative, the Variational Expectation-Maximization (VEM) algorithm is proposed by approximating the E-step. Although VEM is commonly used to estimate communities in different variations of the SBM, it presents challenges when applied to zero-inflated models, as the maximization step lacks a closed-form solution, requiring numerical optimization. To address this, we propose using an Expectation Conditional Maximization (ECM) algorithm in the M-step for parameter inference. In our contribution, we derived closed-form expressions for the parameter estimators in the M-step by introducing a latent variable that indicates whether an observed zero is generated by the structural process or by the Poisson process in the ZIP model. Simulations studies carried out in this work showed that the proposed algorithm performs comparably to other competing algorithms, standing out as the best method for community detection in the case of sparser and more unbalanced networks.

The proposed community detection method assumes that the number of communities is known, as is common in solving this problem. To estimate the number of communities in the network, we propose a method based on the Integrated Completed Likelihood (ICL), originally introduced by \cite{Biernacki2000} for mixture models and extended to the SBM by \cite{daudin2008mixture}.

In this work, we opted for the ZIP model over the hurdle model because it may be more suitable for some real-world applications. For instance, in modeling airport networks, the ZIP model is preferred as it effectively accounts for structural zeros— representing the permanent absence of direct flights between certain airports— and the variability of edge weights, which reflect the number of flights between connected airports over time, that can influenced by external factors such as COVID-19 travel restrictions. In contrast, the hurdle model treats zeros and positive counts separately, which may not fully capture the underlying nature of zeros arising from the variability in flight numbers. In this way, we apply the proposed community detection method to analyze the Brazilian airport network, with the primary goal of comparing the communities identified in the domestic flight networks for the years 2019 and 2020. This comparison aims to evaluate the impact of travel restrictions imposed during the COVID-19 pandemic. The proposed method is well-suited to this network, which exhibits a high presence of non-edges and significant hubs. These characteristics make the degree-corrected model especially useful, allowing us to capture both network sparsity and the varying connectivity strengths of hub airports.

This work is organized as follows. In Section \ref{sec:Model}, we introduce the SBM model based on the ZIP distribution. Section \ref{sec:inference} presents the proposed community detection algorithm based on the likelihood. In Section \ref{sec:icl}, we discussed the model selection problem. Section \ref{sec:simulation} provides a simulation study to evaluate the performance of the proposed method. Section \ref{sec:data} presents the results of the analysis of the Brazilian airport network.  A discussion section is included in Section \ref{sec:discussion}.  The mathematical derivation of the community detection method is detailed in the Appendix.

\section{Model}\label{sec:Model}
   
We consider a directed weighted network model with discrete edge weights in which the $n$ nodes are partitioned into $K$ communities. The latent random vector $\Z_i =(Z_{i1},\dots, Z_{iK})$ represents the community assignment of node $i$ such that $Z_{ia}=1$ indicates that node $i$ belongs to community $a$, and $Z_{ia}=0$ otherwise. We assume that the communities are chosen independently in the network, thus $\Z_1,\dots,\Z_n$ are mutually independent random vectors. The communities assignments are described by a Multinomial distribution, that is, for $i=1,\dots,n$
\begin{equation}\label{def:Z}
    {\Z}_i\sim \mathcal{M}(1, \npi=(\pi_1, \cdots, \pi_K))\,,
\end{equation}
where $\pi_a$ is the probability that a node belongs to community $a$, $a=1,\dots,K$. 

In order to model networks with discrete weights, the adjacency matrix $\A$ contains positive integers values describing the strength of the connection between each pair of nodes. Each entry of the adjacency matrix $A_{ij}$ is conditionally independent given the community assignments ${\Z}_1,\dots,{\Z}_n$. In weighted networks, the presence of zero entries in the adjacency matrix can have two different interpretations: it can represent either the absence of an edge or an edge with zero strength. These interpretations can lead to distinct network analyses, as the absence of an edge implies no interaction between nodes, while a zero weight indicates the presence of an interaction with negligible strength at the time the network was observed.

In this work, we aim to model weighted networks with discrete weights that are sparse, that is, we are considering the case that the zero entries represents the absence of edges. Thus, we model the edge weights as a mixture of a Poisson and a degenerate distribution  at zero, using a Zero-Inflated Poisson (ZIP) distribution to describe the distribution of edge weights. 

 In our case, we also want to capture the heterogeneity of the nodes strength, that is,  the sum of weights of the edges connected to each node of the network. \cite{karrer2011} proposed the Degree-Corrected Stochastic Block Model (DCSBM), an extension of the Stochastic Block Model (SBM) proposed by \cite{holland1983stochastic}, which accounts for the heterogeneity of node degrees.
 However, in order to ensure the identifiability of the model, they imposed a constraint on the parameters that model the nodes' degrees, making them dependent on the community assignments $\Z_1,\dots,\Z_n$. This dependence introduces additional complexity in the estimation process when an Expectation-Maximization (EM) type of algorithm is applied to estimate the communities.
 Following the approach proposed in \cite{riverain2023poisson}, we assume that each node $i$ is associated with parameters $\mu_i$ and $\nu_i$, which model its out-strength and in-strength, respectively. Thus, we define the weights of the edges as

\begin{equation}\label{def:sbm_zip}
          A_{ij}\mid Z_{ia}Z_{jb}=1 \sim \text{ZIP}(p_{ab},\mu_i\nu_j\lambda_{ab})\,,\qquad i\neq j\,
\end{equation}
where $\p\in [0,1]^{K\times K}$ and $\nLambda\in \mathbb{R}^{K \times K}$ are symmetric matrices  and  $\text{ZIP}(p_{ab},\mu_i\nu_j\lambda_{ab})$ corresponds to the ZIP distribution with probability distribution defined on $\mathbb{Z}_{+}$ given by
\begin{equation}\label{def:zip}
  f_{ab}(a_{ij}) = \left\lbrace
  \begin{aligned}
      &p_{ab}+(1-p_{ab})e^{-\mu_i\nu_j\lambda_{ab}} &, \text{ if } a_{ij}=0  \\
      &(1-p_{ab}) \dfrac{(\mu_i\nu_j\lambda_{ab})^{a_{ij}}e^{-\mu_i\nu_j\lambda_{ab}}}{a_{ij}!}  &, \text{ if } a_{ij}\neq 0.
  \end{aligned}
  \right.
\end{equation}\\

The proposed model does not take self-loops into account; therefore, the diagonal elements of the matrix $\A$ are set to zero.
This model can be interpreted as follows: given that nodes $i$ and $j$ belong to communities $a$ and $b$, respectively, with probability $p_{ab}$ there is no edge between $i$ and $j$. If the edge exists, then its weight has a Poisson distribution with mean $\lambda_{ab}$. In this model, the parameters $\p$, which are associated with the existence of edges, and the parameters $\nLambda$, which control the weights and depend on the community of the nodes, allow the model to capture a \textit{local} structure present in each group of nodes. Thus, we can model a group of nodes that is sparser than the other groups in the network by adjusting the parameter $p_{ab}$. On the other hand, this model is also flexible enough to describe a \textit{global} structure in the network, where one parameter $p$ controls the sparsity of the entire network. For global models, we can consider the model described in \eqref{def:sbm_zip} with $p_{ab}=p$, for $a,b=1,\dots,K$. In the remaining parts of the paper we consider the local model \eqref{def:sbm_zip}, but the method proposed can be easily extended to global model by minor modifications of the algorithms.

The in-strength of the node $i$ in the network is computed by summing all the weights of the edges pointing to node $i$ and its expected value is given by 
\begin{equation}
\nu_i\sum_{j\neq i}\mu_j\sum_{a,b=1}^{K} (1-p_{ab})\lambda_{ab}\pi_{a}\pi_{b}.
\end{equation}
In the same way, the out-strenght of node $i$ is computed by summing all the weights of the edges pointing out from node $i$ with expected value given by
\begin{equation}
\mu_i\sum_{j\neq i}\nu_j\sum_{a,b=1}^{K} (1-p_{ab})\lambda_{ab}\pi_{a}\pi_{b}.
\end{equation}

As the expected out and in strengths depend on its associated parameters \( \nmu \) and \( \nnu \), respectively, we refer to the model proposed by \eqref{def:sbm_zip} as the Degree Corrected ZIP (DCZIP) model. To consider the model without the degree correction it is enough to set all entries of $\nnu$ and $\nmu$ equal to one. According to this proposed model, the distribution of $\Z$ is given by
\[\mathbb{P}(\Z\mid\npi)=\prod_{i=1}^{n}\mathbb{P}(\Z_i|\npi)= \prod_{i=1}^{n}\prod_{a=1}^{K} \pi_a^{Z_{ia}}, \]
 and the conditional distribution of $\A$ given $\Z$ is 
\begin{equation}\label{cond_dist}
        \mathbb{P}(\A\mid \Z,\nLambda, \p,\nmu,\nnu)= \prod_{i\neq j}\prod_{a,b=1}^{K} f_{ab}(A_{ij})^{Z_{ia}Z_{jb}}.
\end{equation}

Denoting the model parameters by $\ntheta = (\npi,\nLambda,\p,\nmu,\nnu) $, we can write the complete data log-likelihood by
\begin{equation}\label{eq:vero}
      \log\,\mathbb{P}(\A,\Z\mid \ntheta)= \sum_{i=1}^{n}\sum_{a=1}^{K} Z_{ia}\log(\pi_a) +  \sum_{i\neq j}\sum_{a,b=1}^{K} Z_{ia}Z_{jb}\log f_{ab}(A_{ij})\,.
  \end{equation}
  
\section{Inference}\label{sec:inference}
In this section, we address the problem of detecting the communities associated with each node and estimating the model parameters \( \boldsymbol{\theta} \) given an observed network \( \mathbf{A} \) generated from the model defined in \eqref{def:sbm_zip}. It is well known that the traditional Expectation-Maximization (EM) algorithm, widely used for mixture models, is intractable for estimating the communities in variants of the stochastic block model \cite{daudin2008mixture}. This is because in the E-step of the EM algorithm, it is necessary to compute the conditional distribution of the latent communities \( \mathbf{Z} \) given the data \( \mathbf{A} \), \( \P(\mathbf{Z} \mid \mathbf{A}, \boldsymbol{\theta}) \), which is intractable for networks due to the strong dependence on the network edge structure. To overcome this issue, we approximate the conditional distribution \( \P(\mathbf{Z} \mid \mathbf{A}, \boldsymbol{\theta}) \) with a distribution \( Q(\mathbf{Z}) \).

Given a distribution $Q(\Z)$, the log-likelihood, also called evidence, is decomposed into two terms
\begin{equation}
  \log \mathbb{P}(\A \mid \ntheta)= \KL(Q(\cdot) \parallel \mathbb{P} (\cdot\mid \A, \ntheta) ) + \mathfrak{L}(Q(\cdot), \ntheta)\,,
  \label{dec}
\end{equation}
where
\begin{equation}\label{eq:KL}
    \KL(Q(\cdot) \parallel \mathbb{P} (\cdot\mid \A, \ntheta) ) = -\sum\limits_{\Z} Q(\Z)\log\left(\dfrac{\mathbb{P}(\Z\mid\A, \ntheta)}{Q(\Z)}\right)\,,
\end{equation}
and
\begin{equation}\label{elbo}
 \begin{split}
    \mathfrak{L}(Q(\cdot),\ntheta) = & \sum_{\Z} Q(\Z) \log \left(\dfrac{\mathbb{P}(\A,\Z\mid \ntheta)}{Q(\Z)}\right) \\
     &=  \mathbb{E}_Q[\log \mathbb{P}(\A,\Z\mid  \ntheta)] - \E_Q[\log Q(\Z)].
     \end{split}
\end{equation}

The first term in the decomposition given by \eqref{eq:KL} is the Kullback-Leibler (KL) divergence between the proposed distribution \( Q(\mathbf{Z}) \) and \( \mathbb{P}(\mathbf{Z} \mid \A, \boldsymbol{\theta}) \). Since 
 the KL divergence is positive and the likelihood of the data is fixed, the second term in \eqref{elbo} gives a lower bound for the likelihood, known as the evidence lower bound (ELBO).

To ensure that \( Q(\mathbf{Z}) \) provides a good approximation of \( \mathbb{P}(\mathbf{Z} \mid \mathbf{A}, \boldsymbol{\theta}) \), given that the log-likelihood is fixed, it is necessary to maximize the ELBO, \( \mathfrak{L}(Q(\cdot), \boldsymbol{\theta}) \), with respect to \( Q(\mathbf{Z}) \). This maximization aims to minimize the KL divergence between these two distributions. To achieve a tractable method for maximizing the ELBO, we propose a variational distribution \( Q(\mathbf{Z}) \) that factorizes as
\begin{equation}\label{eq:Q_dist}
    Q(\Z)= \prod^{n}_{i=1}Q(\Z_i)= \prod_{i=1}^{n}\prod_{a=1}^{K} \tau_{ia}^{Z_{ia}}\,,
\end{equation}
where $\nt_i=(\tau_{i1},\dots,\tau_{iK})$ and $\tau_{ia}$ is the probability of node $i$ belongs to community $a$, under the distribution $Q(\cdot)$. 

In this way, the Variational EM agorithm (VEM) is divided in two steps. In the E-step, we find the approximated distribution $Q(\Z)$ in the class of distributions given by the factorization \eqref{eq:Q_dist}  that minimizes the KL divergence, that is, we maximize the ELBO, $\mathfrak{L}(Q(\cdot), \ntheta)$, with respect to $Q(\Z)$ considering that the parameters $\ntheta$ are fixed. In the M-step, we estimate the parameters $\ntheta$  by maximizing the ELBO with respect to these parameters considering that $Q(\Z)$ is fixed. Notice that for fixed $Q(\Z)$, maximizing the ELBO with respect to the parameters is equivalent to maximize the first term in \eqref{elbo}. Thus, at iteration step $t$, the EM-Variational method is given by
\begin{equation}\label{eq:VEM}
    \begin{array}{cc}
       \mbox{E-step:}  &  Q^{(t)}= \arg\max\limits_{Q'}\mathfrak{L}(Q'(\cdot),\ntheta^{(t-1)}) \\[0.5cm]
       \mbox{M-step:} & \ntheta^{(t)}= \arg\max\limits_{\ntheta} \mathfrak{L}(Q^{(t)}(\cdot),\ntheta) \\[0.2cm]
       &\hspace{2.3cm}= \arg\max\limits_{\ntheta} \mathbb{E}_{Q^{(t)}}[\log \mathbb{P}(\A,\Z\mid \ntheta)]\,.
    \end{array}
\end{equation}

In the E-step, the optimal variational parameters $\hat\nt_i$, $i=1,\dots,n$, are obtained as the solution of
\begin{equation}\label{fixed_tau}
    \hat\tau_{ia}   \propto \pi_{a}\prod_{\substack{j=1\\j\neq i}}^n \prod_{b=1}^{K}f_{ab}(A_{ij})^{\hat\tau_{jb}}\,.
\end{equation}
This solution can be obtained by a fixed-point algorithm. Details about the calculations and the fixed-point algorithm are described in Appendix \ref{ape:e-step}.

In the M-step, the updates of the parameters $\pi_a$, $a=1,\dots,K$, can be obtained explicitly through maximizing the ELBO, and are given by
\begin{equation}\label{eq:pi_hat}
    \hat{\pi}_a=\frac{1}{n}\sum_{i=1}^{n}\tau_{ia}.
\end{equation}

Due to the mixture distribution of the ZIP given in \eqref{def:zip}, the update of the parameters $\nLambda, \p, \nmu$ and $\nnu$ in the M-step involves iterative methods based on numerical optimization techniques. In this work, we propose to apply an Expectation Conditional Maximization (ECM) algorithm to estimate these parameters \citep{Meng_ECM1993}. The main idea is to introduce a latent variable that indicates whether a zero (representing the non-existence of an edge between a pair of nodes) is generated by the structural process (Bernoulli part) or by the Poisson process. We first update the parameters $\boldsymbol{p}$, followed by the update of $\boldsymbol{\Lambda}$ given $\boldsymbol{\mu}$ and $\boldsymbol{\nu}$. Then, we update $\boldsymbol{\mu}$ given $\boldsymbol{\Lambda}$ and $\boldsymbol{\nu}$, and finally update $\boldsymbol{\nu}$ given $\boldsymbol{\Lambda}$ and $\boldsymbol{\mu}$. The updates of the parameters at iteration step $s$ are given by
\begin{equation}\label{eq:p_hat}
\hat{p}^{(s)}_{ab}=  \dfrac{\sum_{\substack{i\neq j}}\tau_{ia}\tau_{jb} \alpha_{ij}^{ab,(s)}}{\sum_{\substack{i\neq j}}\tau_{ia}\tau_{jb}},
\end{equation}
\begin{equation}\label{eq:lambda_hat}
    \hat{\lambda}^{(s)}_{ab}=\frac{\sum_{\substack{i\neq j}} \tau_{ia}\tau_{jb} a_{ij}\left ( 1-\alpha_{ij}^{ab,(s)}\right) }{\sum_{\substack{i\neq j}}\tau_{ia}\tau_{jb}\hat\mu_i^{(s-1)}\hat\nu_j^{(s-1)}\left ( 1-\alpha_{ij}^{ab,(s)} \right )},
\end{equation}
\begin{equation}\label{eq:mu_hat}
    \hat{\mu_i}^{(s)}=\frac{\sum_{j=1}^n a_{ij} }{\sum_{j\neq i}\sum_{a,b=1}^K\tau_{ia}\tau_{jb}\hat\nu_j^{(s-1)}\hat\lambda_{ab}^{(s)}\left ( 1-\alpha_{ij}^{ab,(s)} \right )},
\end{equation}
\begin{equation}\label{eq:nu_hat}
    \hat{\nu_j}^{(s)}=\frac{\sum_{i=1}^n a_{ij} }{\sum_{i\neq j}\sum_{a,b=1}^K\tau_{ia}\tau_{jb}\hat\mu_i^{(s)}\hat\lambda_{ab}^{(s)}\left ( 1-\alpha_{ij}^{ab,(s)} \right )}\,.
\end{equation}
where the quantity $\alpha_{ij}^{ab,(s)}$ represents the probability that an observed zero between nodes $i$ and $j$ in communities $a$ and $b$ comes from the structural process (Bernoulli part). The exact expression for this quantity is given in \eqref{eq:passo_em_alpha}. Appendix \ref{sec:M-step} provides a detailed description of the computations used to derive the update formulas.  The VEM algorithm proposed for the DCZIP is described in Algorithm \ref{alg:variational}.

The initial partition $\Z^0$ can have a significant impact on the outputs of the algorithm, as is common with iterative methods. To initialize the algorithm we propose to apply k-means algorithm to the rows of the weighted adjacency matrix $\A$ to obtain a initial partition of the nodes. As initial values for $\nnu$ and $\nmu$ we propose starting with no degree corrections, that is, setting all entries of these vectors equal to one. An alternative way to initialize $\nnu$ and $\nmu$ is by using the average out-strength and in-strength.

\begin{algorithm}
\caption{Variational Expectation-Maximization (VEM) for the DCZIP}
\label{alg:variational}
\begin{algorithmic}[1]
\Statex \textbf{Input:}
\Statex \hspace{\algorithmicindent} - network: $\A$ 
\Statex \hspace{\algorithmicindent} - number of communities: $K$
\Statex \hspace{\algorithmicindent} - initial partition: $\Z^0$
\Statex \hspace{\algorithmicindent} - stopping criterion: $\epsilon$
\Statex \textbf{Output:} 
\Statex \hspace{\algorithmicindent} - vector of estimated communities: $\hat{\Z}_i =(\hat Z_{i1},\dots, \hat Z_{iK})$ 
\Statex \hspace{\algorithmicindent} - estimated parameters of the model: $\hat{\ntheta}$
\Statex \textbf{Initialization:} 
\Statex \hspace{\algorithmicindent} - Set $t=0$
\Statex \hspace{\algorithmicindent} - Set $\tau^{(t)}_{ia}=1$, if $Z^0_{ia}=1$, $1\leq i \leq n$, $1\leq a\leq K$
\Statex \hspace{\algorithmicindent} - Set $\mu^{(t)}_{i}=1$ and $\nu^{(t)}_i=1$, $1\leq i \leq n$
\Statex \hspace{\algorithmicindent} - Set $\npi^{(t)}$, $\p^{(t)}$, $\nLambda^{(t)}$ 
according to \eqref{eq:pi_hat}, \eqref{eq:p_hat} and \eqref{eq:lambda_hat}
\Statex \hspace{\algorithmicindent} - Set $\mathfrak{L}^{(t)}=0$
\Statex  \hspace{\algorithmicindent}\textbf{repeat}
\Statex  \hspace{\algorithmicindent}\hspace{\algorithmicindent} increment $t$
\Statex  \hspace{\algorithmicindent}\hspace{\algorithmicindent} \textbf{E-step:} Update $\nt^{(t)}$ according to \eqref{fixed_tau}
\Statex  \hspace{\algorithmicindent}\hspace{\algorithmicindent} \textbf{repeat}
\Statex  \hspace{\algorithmicindent} \hspace{\algorithmicindent}\hspace{\algorithmicindent} \textbf{M-step:} Update $\npi^{(t)}$ according to \eqref{eq:pi_hat} 
\Statex  \hspace{3.5cm} Update $\npi^{(t)}$, $\p^{(t)}$, $\nLambda^{(t)}$ 
according to \eqref{eq:pi_hat}, \eqref{eq:p_hat} and \eqref{eq:lambda_hat}
\Statex \hspace{\algorithmicindent}\hspace{\algorithmicindent} \textbf{until} convergence of the parameters
\Statex \hspace{\algorithmicindent} \textbf{until}  $\left |  \mathfrak{L}^{(t)} - \mathfrak{L}^{(t-1)} \right |< \epsilon$
\Statex \hspace{\algorithmicindent} Define $a=\arg\max\{\tau_{ik}: 1 \leq k \leq K\}$ and set $\hat{Z}_{ia}=1$, $i=1,\dots,n$
\Statex \textbf{Return:} $\hat{\Z}$, $ \ntheta^{(t)}$
\end{algorithmic}
\end{algorithm}

\section{Model Selection}

The VEM algorithm proposed in Algorithm \ref{alg:variational} depends on the number of communities $K$, which is assumed to be known at this point. Typically, this parameter can be estimated through model selection methods, such as using information criteria like BIC or ICL, before applying the algorithm.
The Integrated Completed Likelihood (ICL) criterion was proposed by \cite{Biernacki2000} to select the number of mixture components in a finite mixture model. The ICL criterion addresses the trade-off between model complexity and goodness of fit by penalizing an approximation of the complete data likelihood. \cite{daudin2008mixture} adapted the ICL criterion to estimate the number of communities in the Stochastic Block Model. We derive the ICL for the proposed DCZIP model and obtain
\begin{equation}
    \text{ICL}_k(\A) = \max\limits_{\ntheta}  \log \mathbb{P}(\A, \hat{\Z} \mid \ntheta) - \left(\frac{k(k+1)}{2} + n \right) \log\left({n(n-1)}\right)-\frac{k-1}{2}\log n\,.
\end{equation}

It is also possible to obtain the ICL for the model without the degree correction, that is, when $\mu_i$ and $\nu_i$ are assumed to be known and equal to one for all nodes of the network. Thus, the ICL is given by
\begin{equation}
    \text{ICL}_k(\A) = \max\limits_{\npi,\Lambda,\p}  \log \mathbb{P}(\A, \hat{\Z} \mid \npi,\Lambda,\p) - \frac{k(k+1)}{2}\log\left(n(n-1)\right)-\frac{k-1}{2}\log n\,.
\end{equation}

Thus, the estimator of the number of communities based on the ICL is given by

\begin{equation}
    \hat k ( \A) = \arg\max_{k} \text{ICL}_k(\A)\,.
\end{equation}

It should be noted that there are currently no theoretical results on the convergence of the ICL procedure, neither in simple mixture models nor in the case of the SBM. However, the simulation study carried out in Section \ref{sec:simulation} showed that the method performances well in recovering the true number of communities as the number of nodes in the network increases.

\section{Simulation Study}\label{sec:simulation}

In this section we study the performance of the VEM algorithm on synthetic data. We compare the proposed VEM algorithm for discrete weighted networks to four other methods, the Spectral Clustering \citep{Lei_SC_2015} and Spherical Spectral Clustering \citep{qin2013regularized} developed for binary networks are applied directly to the weighted matrix, the Louvain method based on the modularity optimization algorithm applied for weighted networks \cite{blondel2008fast} and we apply the VEM algorithm without considering the degree correction in the ZIP model. The VEM algorithm is initialized applying k-means algorithm to the rows of the weighted network.

We simulate networks from the model \eqref{def:sbm_zip} and the performance of the proposed method and others is computed by means of the Normalized Mutual Information (NMI) averaged over 100 replications. The NMI is a similarity measure commonly used to compare clusters by measuring the mutual information shared by two partitions. The NMI reaches its minimum value of 0 when the clustering is completely random and unrelated to class membership. On the other hand, it reaches its maximum value of 1 when the clustering perfectly matches the true classes.

\begin{figure}[H]
     \centering
     \begin{subfigure}[b]{0.5\textwidth}
         \centering
         \includegraphics[scale=0.5]{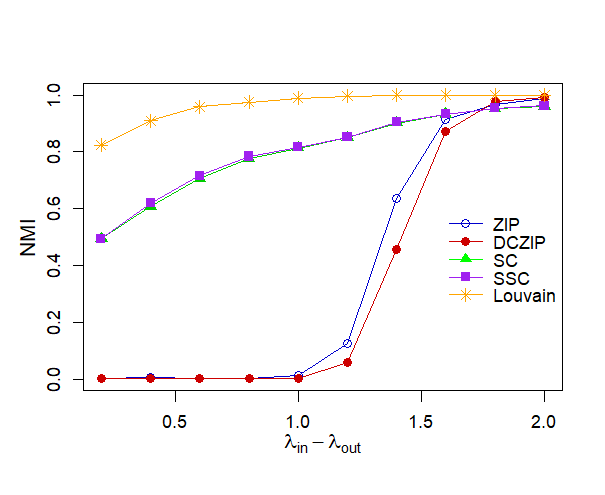}
         \caption{balanced communities: $\npi=(0.5,0.5)$.}
         \label{fig:NMI_lambda_a}
     \end{subfigure}\hfill
     \begin{subfigure}[b]{0.5\textwidth}
         \centering
         \includegraphics[scale=0.5]{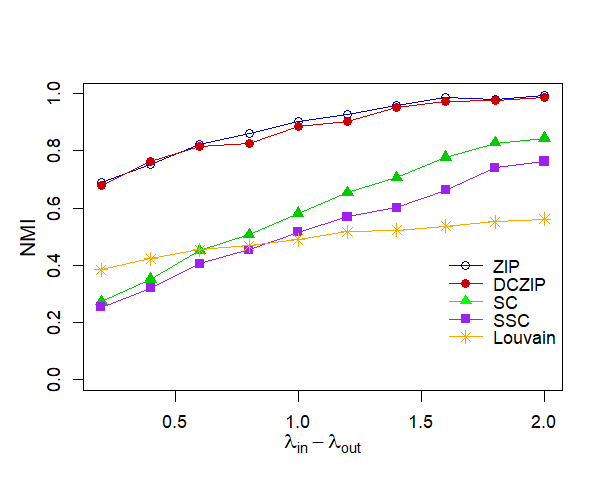}
         \caption{unbalanced communities: $\npi=(0.7,0.3)$.}
         \label{fig:NMI_lambda_b}
     \end{subfigure}
     \caption{Mean of NMI between estimated and true communities membership over 100 simulated networks with $n=100$. The networks are sampled from the ZIP model without degree correction and with mean weights $\lambda_{\text{out}}$ (between groups) and $\lambda_{\text{in}}$ (within groups).}
     \label{fig:NMI_lambda}
\end{figure}

First, we analyze the impact of parameter $\nLambda$ in the performance of the algorithm. To do so, we consider networks with $K=2$, $n=100$ and fixed the off-diagonal of $\nLambda$, $\lambda_{\text{out}}$, equal to $5$, that is, the average of the weights of existing edges between communities. To control the local sparsity of the network, we fixed the off-diagonal elements of matrix $\p$ to be equal to 0.7 (between communities) and the diagonal elements to be equal to 0.5 (within communities). Thus, we consider networks with fewer edges between groups. We set the out and in strength, $\nmu$ and $\nnu$, equal to 1 for all  nodes. In this setting, the nodes have the same expected degree. In Figure \ref{fig:NMI_lambda_a}, for balanced communities, we observe that with a small difference in $\lambda_{\text{in}} - \lambda_{\text{out}}$, Spectral Clustering (SC), Spherical Spectral Clustering (SSC), and the Louvain algorithm outperform the VEM algorithm with degree correction (DCZIP) and without degree correction (ZIP). However, for unbalanced communities, as shown in Figure \ref{fig:NMI_lambda_b}, the proposed VEM algorithm performs better than the other methods. We also observe that the VEM algorithm proposed for the ZIP model performs similarly with and without degree correction, as in this setting, the nodes have the same expected weights.

\begin{figure}[H]
     \centering
     \begin{subfigure}[b]{0.5\textwidth}
         \centering
         \includegraphics[scale=0.5]{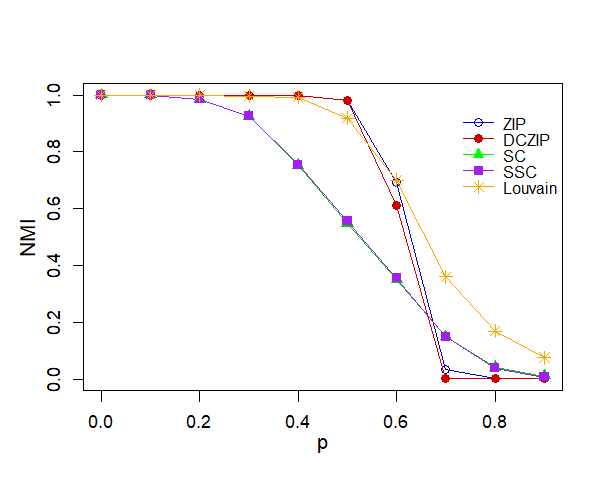}
         \caption{balanced communities: $\npi=(0.5,0.5)$.}
          \label{fig:NMI_p_a}
     \end{subfigure}\hfill
     \begin{subfigure}[b]{0.5\textwidth}
         \centering
         \includegraphics[scale=0.5]{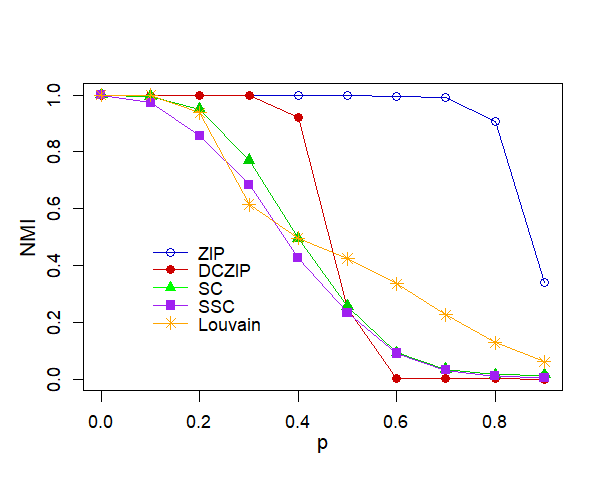}
         \caption{unbalanced communities: $\npi=(0.7,0.3)$.}
         \label{fig:NMI_p_b}
     \end{subfigure}
     \caption{Mean of NMI between estimated and true communities membership over 100 simulated networks with $n=100$. The networks are sampled from the ZIP model without degree correction and with mean weights $\lambda_{\text{out}}=5$ (between groups) and $\lambda_{\text{in}}=8$ (within groups). The parameter $p$ controls the global sparsity of the networks.}
     \label{fig:NMI_p}
\end{figure}

To study the impact of sparsity in order to correctly recover the communities, we set $\lambda_{\text{in}}=8$ and $\lambda_{\text{out}}=5$. We set the entries of $\p$ equal to $p$ such that $p$ controls the global sparsity of the network, which represents the probability of an edge existing between any pair of nodes. We varied the values of parameter $p$ from 0 to 0.9. We observe, as expected, that in the balanced case (Figure \ref{fig:NMI_p_a}), the method's performance decreases as the sparsity of the networks increases (as $p$ increases). However, in the case of unbalanced communities, the VEM method without degree correction of the nodes outperforms the other methods, followed by the VEM method with degree correction, Figure \ref{fig:NMI_p_b}.

\begin{figure}[H]
     \centering
     \begin{subfigure}[b]{0.5\textwidth}
         \centering
         \includegraphics[scale=0.5]{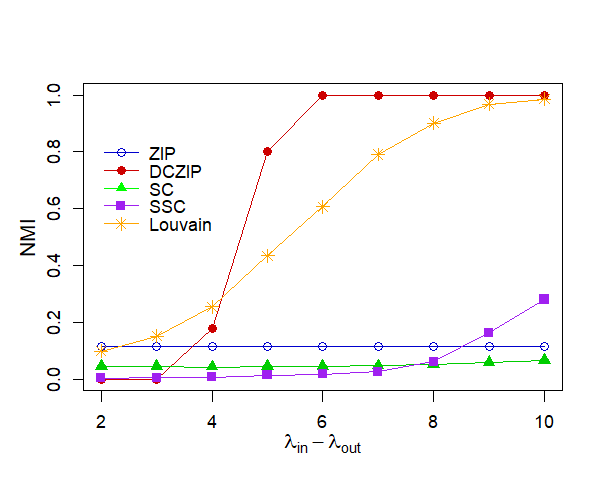}
         \caption{balanced communities: $\npi=(0.5,0.5)$.}
          \label{fig:NMI_lambda_degree_a}
     \end{subfigure}\hfill
     \begin{subfigure}[b]{0.5\textwidth}
         \centering
         \includegraphics[scale=0.5]{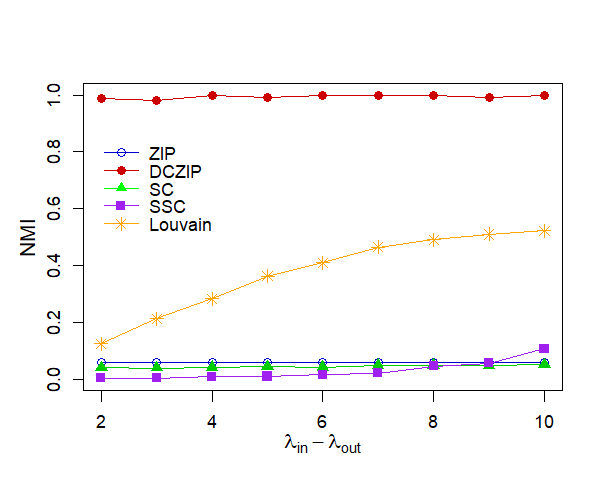}
         \caption{unbalanced communities: $\npi=(0.7,0.3)$.}
         \label{fig:NMI_lambda_degree_b}
     \end{subfigure}
     \caption{Mean of NMI between estimated and true communities membership over 100 simulated networks with $n=100$. The networks are sampled from the ZIP model with degree correction and with mean weights $\lambda_{\text{out}}$ (between groups) and $\lambda_{\text{in}}$ (within groups).}
     \label{fig:NMI_lambda_degree}
\end{figure}

\begin{figure}[H]
     \centering
     \begin{subfigure}[b]{0.5\textwidth}
         \centering
         \includegraphics[scale=0.45]{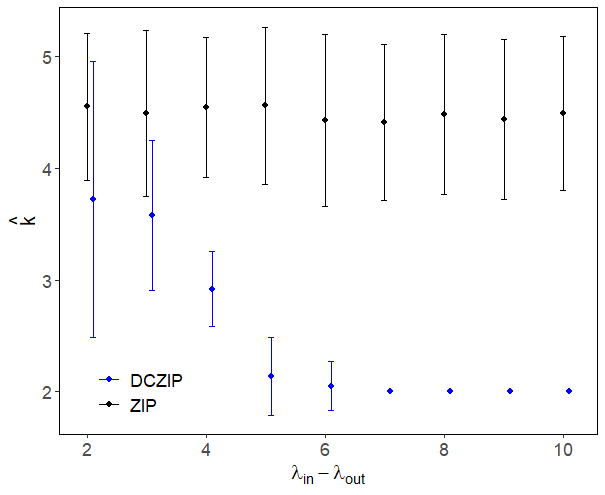}
         \caption{$n=100$.}
     \end{subfigure}\hfill
     \begin{subfigure}[b]{0.5\textwidth}
         \centering
         \includegraphics[scale=0.45]{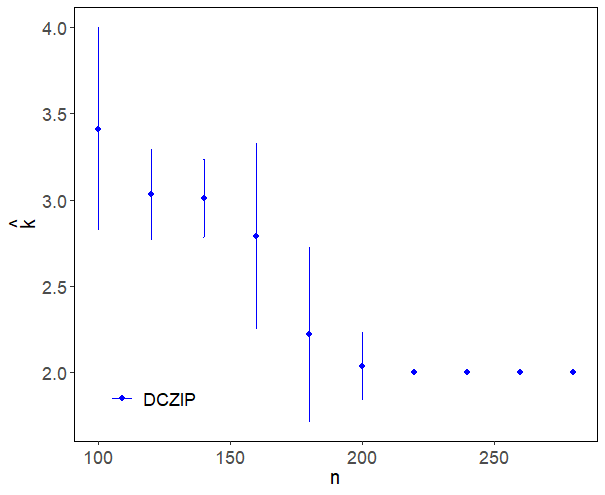}
         \caption{ $\lambda_{\text{in}}-\lambda_{\text{out}}=3$.}
     \end{subfigure}
     \caption{The mean and the one standard deviation error bars for the estimated number of communities over 100 simulated networks with balanced communities. The networks are sampled from the ZIP model with degree correction and with mean weights $\lambda_{\text{out}}$ (between groups) and $\lambda_{\text{in}}$ (within groups).}
     \label{fig:ICL_diff}
\end{figure}

To analyze networks with hubs, we use the same setting as before but modify the out-strength of 15\% of the nodes in community 1 and the in-strength of 15\% of the nodes in community 2 by 8. In Figure \ref{fig:NMI_lambda_degree}, we vary the value of the diagonal of $\nLambda$, $\lambda_{\text{in}}$, and observe that the proposed VEM algorithm with degree correction performs better than the others in both balanced and unbalanced cases. This may occur because the other algorithms tend to cluster nodes with high strength together in the same community, whereas the VEM method is designed to detect this difference. In this setting, we also study the ICL method to recover the number of communities, as shown in Figure \ref{fig:ICL_diff}. Fitting the ZIP model to the sampled networks overestimates the true number of communities ($K=2$). The criterion performs well when the corrected ZIP model is applied, particularly as the difference between the in-strength and out-strength grows and as the number of nodes in the network increases.

\begin{figure}[H]
     \centering
     \begin{subfigure}[b]{0.5\textwidth}
         \centering
         \includegraphics[scale=0.5]{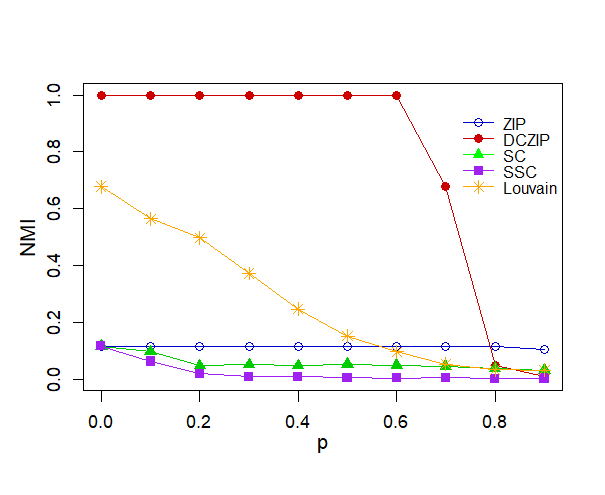}
         \caption{balanced communities: $\npi=(0.5,0.5)$.}
          \label{fig:NMI_p_degree_a}
     \end{subfigure}\hfill
     \begin{subfigure}[b]{0.5\textwidth}
         \centering
         \includegraphics[scale=0.5]{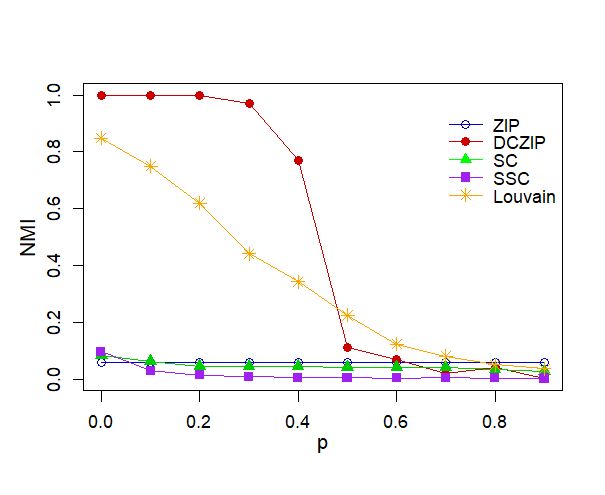}
         \caption{unbalanced communities: $\npi=(0.7,0.3)$.}
         \label{fig:NMI_p_degree_b}
     \end{subfigure}
     \caption{Mean of NMI between estimated and true communities membership over 100 simulated networks with $n=100$. The networks are sampled from the ZIP model with degree correction and with mean weights $\lambda_{\text{out}}=5$ (between groups) and $\lambda_{\text{in}}=11$ (within groups). The parameter $p$ controls the global sparsity of the networks.}
     \label{fig:NMI_p_degree}
\end{figure}

To study the effect of sparsity in the presence of hubs, we set $\lambda_{\text{in}}=11$ and $\lambda_{\text{out}}=5$ and varied the network sparsity controlled by $p$ from 0 to 0.9. In the presence of hubs, Figure \ref{fig:NMI_p_degree}, we observed that the proposed VEM algorithm with degree correction outperformed the other methods in both balanced and unbalanced cases. As expected, the performance  of this method decreased as the sparsity of the network increased.

\section{Analysis of Airport network}\label{sec:data}

We apply the community detection method to analyze the Brazilian Airport Network. The data is publicly available on the National Civil Aviation Agency of Brazil website\footnote{Data available at https://www.gov.br/anac/pt-br}. We consider domestic flights from the years 2019 and 2020, where the nodes of the network represent the airports, and the edge connecting nodes $i$ and $j$ represents the number of flights from airport $i$ to airport $j$. To ensure that the two networks have the same vertex set, we consider only airports that had at least one flight in both 2019 and 2020. As a result, each network has $n=198$ nodes, with densities of 0.0319 and 0.0317 in 2019 and 2020, respectively. The mean number of routes for airports is 6.35 in 2019 and 6.31 in 2020. Among the five airports with the highest number of routes, three are located in São Paulo state (Campinas, Guarulhos, and São Paulo), one is located in Minas Gerais state (Confins), and the last one is in the capital of the country, Brasília. These airports also have the highest in-strength.  It is worth noting that although the networks have similar densities and mean numbers of routes, the mean edge strength in 2019 is 3865.869, which decreases by almost half to 1904.778 in 2020. This difference is likely due to the impact of transportation restrictions caused by the COVID-19 pandemic, which started in Brazil in March 2020.

We started using the ICL criterion to select the number of groups resulting in 6 communities. We then applied the VEM algorithm without degree correction (ZIP), initialized using the k-means algorithm on the rows of the weighted network, to both networks for the years 2019 and 2020. After that, we used the resulting communities as the initial partition in the VEM algorithm with degree corrections (DCZIP). By Table \ref{tab:NMI_airports}, we observe that the NMI values are significantly lower, suggesting variations in community structure over the years and between models.

\begin{table}[ht]
\centering
\begin{tabular}{cc|cc|cc}
  \hline
  & & \multicolumn{2}{c}{ZIP} & \multicolumn{2}{|c}{DCZIP} \\  
 & & 2019 & 2020 & 2019 & 2020 \\ 
  \hline
\multirow{2}{*}{ZIP} & 2019 & 1.00 & 0.14 & 0.08 & 0.07 \\ 
  & 2020 &  & 1.00 & 0.06 & 0.08 \\ \hline
  DCZIP & 2019 &  &  & 1.00 & 0.12 \\ 
   \hline
\end{tabular}
\caption{ NMI values comparing the partitions obtained by the ZIP and DCZIP models for the years 2019 and 2020.} 
\label{tab:NMI_airports}
\end{table}

\begin{figure}[ht]
    \centering
    \begin{subfigure}[b]{\textwidth}
        \centering
        \includegraphics[scale=0.5]{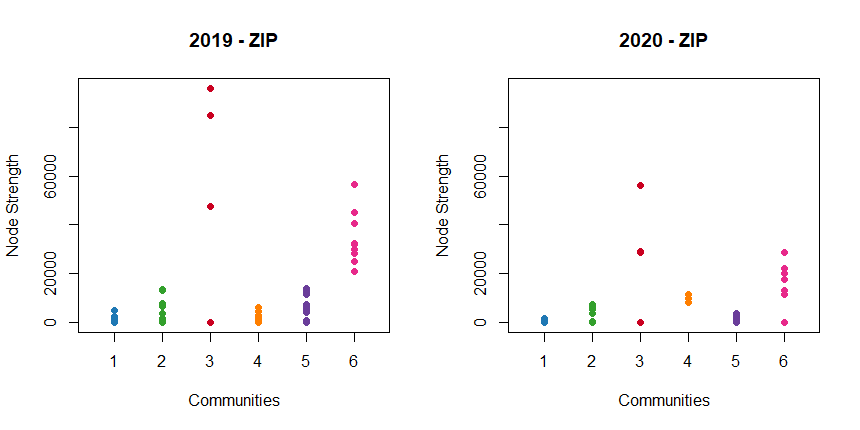}
    \end{subfigure}
    \vspace{0.5cm}  
    \begin{subfigure}[b]{\textwidth}
        \centering
        \includegraphics[scale=0.5]{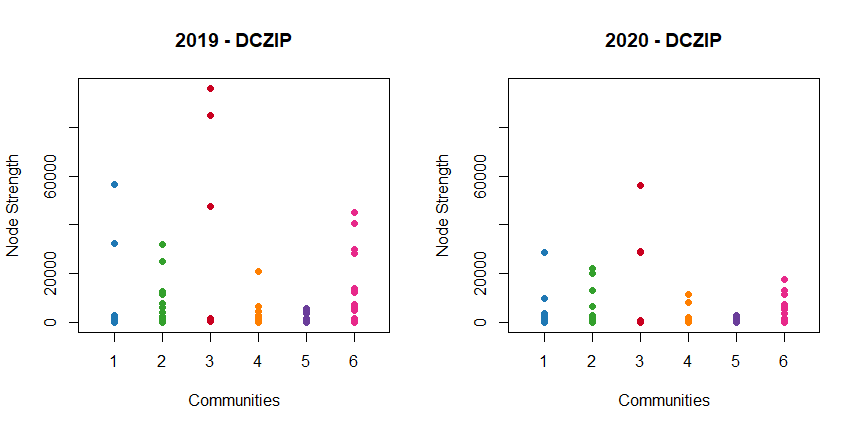}
    \end{subfigure}
    
    \caption{Node in-strength distribution across communities.}
    \label{fig:strength_groups}
\end{figure}

Figure \ref{fig:strength_groups} shows the distribution of node in-strength values across the communities obtained from the ZIP and DCZIP models. We observe that the ZIP model, as expected, tends to group nodes with high in-strength within the same community, while the DCZIP model tends to distribute nodes with high in-strength across different communities due to the correction factor applied to each node. This effect is clearly shown in Table \ref{tab:mean_strength}, which indicates that the average node in-strength varies significantly across groups and years, with values being more distributed across communities in the DCZIP model.

\begin{table}[ht]
\centering
\begin{tabular}{c|cc|cc||cc|cc}
  \hline
    & \multicolumn{4}{c|}{ZIP} & \multicolumn{4}{|c}{DCZIP} \\  
    \cline{2-9}
    & \multicolumn{2}{c|}{2019} & \multicolumn{2}{c|}{2020} & \multicolumn{2}{c|}{2019} & \multicolumn{2}{c}{2020} \\
    \cline{2-9}
  Groups & n & strength & n & strength & n & strength & n & strength \\ 
  \hline
1 & 113 & 330.86 & 113 & 157.54 & 100 & 1095.10 & 126 & 545.27 \\ 
  2 & 16 & 3090.69 & 15 & 2749.00 & 22 & 4934.32 & 15 & 4981.13 \\ 
  3 & 9 & 25447.33 & 11 & 10374.18 & 9 & 26013.89 & 7 & 16520.86 \\ 
  4 & 20 & 1840.35 & 3 & 9802.00 & 33 & 1704.06 & 17 & 1778.53 \\ 
  5 & 31 & 3296.58 & 48 & 994.02 & 17 & 1420.29 & 21 & 629.43 \\ 
  6 & 9 & 34508.56 & 8 & 15859.25 & 17 & 13698.41 & 12 & 6218.83 \\ 
   \hline
\end{tabular}
\caption{Number of nodes (n) and average node in-strength for each group in the ZIP and DCZIP models.} 
\label{tab:mean_strength}
\end{table}

\begin{figure}[!ht]
    \centering
    \begin{subfigure}[b]{\textwidth}
        \centering
        \includegraphics[scale=0.5]{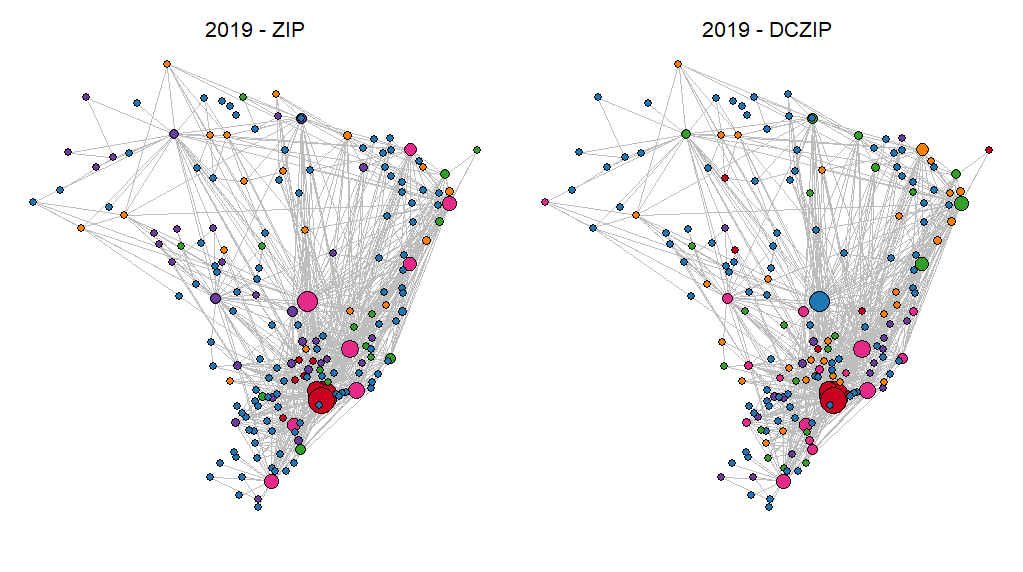}
    \end{subfigure}
    \vspace{0.5cm}  
    \begin{subfigure}[b]{\textwidth}
        \centering
        \includegraphics[scale=0.5]{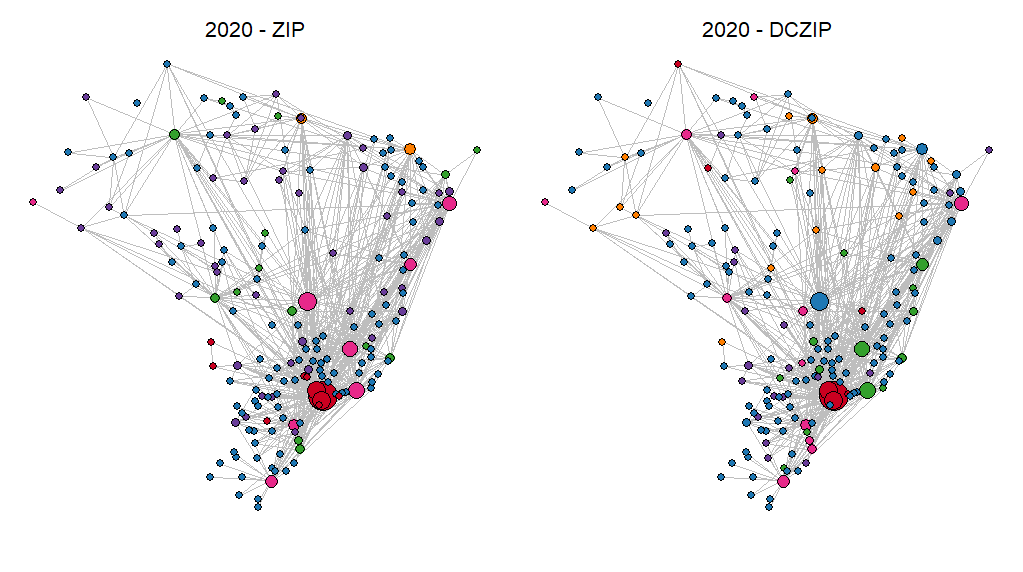}
    \end{subfigure}
    
    \caption{Communities estimated by the VEM algorithm, with node sizes proportional to node in-strength.}
    \label{fig:map_communities}
\end{figure}

Figure \ref{fig:map_communities} presents maps of Brazilian airports, with colors indicating the communities estimated using the ZIP and DCZIP models. The sizes of the nodes are proportional to the node in-strength, illustrating the relative significance of each airport within the estimated communities. Nodes with high in-strength are allocated in the same community by the ZIP model without degree correction. In particular, we can highlight that 7 and 10 airports located in the capitals of the states are clustered together in group 1 for the years 2019 and 2020, respectively. Comparing both groups, we observe an overlap of 5 airports. Notably, the airport in the capital of Brazil, Brasília, is located in group 1 with an in-strength of 56,611 flights in 2019 and 28,714 flights in 2020. 

\begin{figure}[!h]
    \centering
    \includegraphics[scale=0.52]{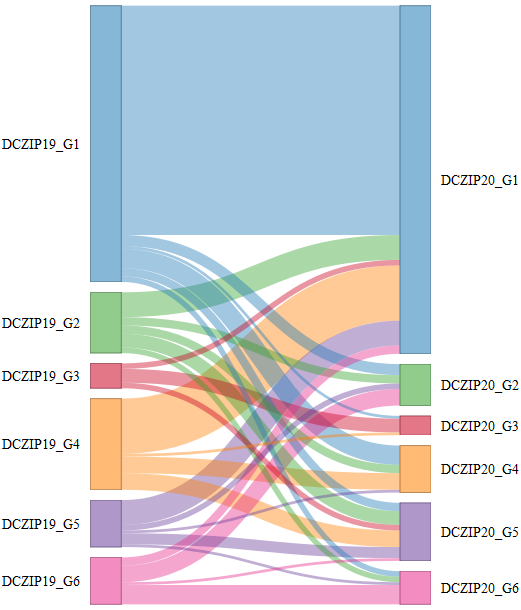}
    \caption{Sankey diagram for comparing the estimated communities in the 2019 and 2020 airport networks.}
    \label{fig:sankey_plot}
\end{figure}

We compare the communities identified using degree correction for both 2019 and 2020 through a Sankey plot (Figure \ref{fig:sankey_plot}), which visually represents the flow and connection of community memberships across the two years. Notably, a significant proportion of airports (60\%) from group 4, which had a moderate mean in-strength in 2019, shifted to group 1 in 2020, which has a lower mean in-strength. For instance, the airport in Fortaleza, the capital and most populous city of Ceará, saw its in-strength decrease from 20,792 in 2019 to 9,792 in 2020, possibly due to its role as a tourist destination, which may have been impacted by the pandemic. 

Community 3 in both years contains a small number of airports with a high average in-strength, and four of these airports are located in São Paulo state (Campinas, São Paulo, Guarulhos and São José dos Campos). Despite a decrease in in-strength, these airports remained in the same group across both years, likely because they are in economically significant regions rather than tourist destinations, which may have made them less susceptible to the impacts of the pandemic. Fernando de Noronha is a Brazilian archipelago known as a tourist destination, and its airport belonged to group 3 in 2019 with an in-strength of 1,708, but changed to group 5 due to a decrease in in-strength to 594.

\section{Discussion}\label{sec:discussion}

In this work, we propose an extension of the SBM model for weighted networks that incorporates both network sparsity and edge weights. The model incorporates degree corrections of the nodes to capture variations in node connectivity. The edge weights follow a zero-inflated Poisson distribution, which assumes two processes: one representing structural zeros and the other capturing the variability of the strength of existing connections. This approach is useful for modeling airport networks because it allows us to handle the fact that some routes may have no flights at all (structural zeros) while capturing varying flight frequencies on active routes. 
We propose a variational EM algorithm to estimate the community membership and the model parameters.  Additionally, we proposed a model selection criterion based on the ICL. We derived closed-form expressions for the parameter estimators obtained using the ECM algorithm. 

In the simulation analysis, we observed that the VEM algorithm performs similarly to other methods for balanced networks, but it outperforms the others in networks with unbalanced communities. Although the algorithm  depends on the choice of initial community membership, which is common in iterative models, we found that it produces reliable estimates and converges well in different simulation scenarios.

We apply the proposed algorithm to recover the Brazilian airport network’s community structure, effectively capturing its sparse connections and hubs, for which the DCZIP model is particularly useful. While routes remained similar between 2019 and 2020, flight frequency shifted substantially in 2020. Notably, communities differed between the two years, especially for airports in tourist regions, likely due to COVID-19 restrictions.

Finally, several extensions of the zero-inflated SBM remain to be studied, such as incorporating mixed membership, analyzing time-evolving networks, and modeling multilayer networks with degree correction. Additionally, it would be interesting to investigate more complex models that take node covariates into account.

\section{Acknowledgments}

This research was supported by the São Paulo Research Foundation (FAPESP)
grant 2023/05857-9 to A.C.

\appendix\label{sec:app}
\section{Variational E-step}\label{ape:e-step}

In this section, we derive all the calculations involved in the E-step of the variational EM algorithm. In E-step, we maximize the ELBO using the variational result which states that the optimal approximated distribution is such that

\begin{equation}
        \log Q(\Z_i) = \mathbb{E}_{\Z_j \mid j \neq i}[\log \mathbb{P}(\A,\Z\mid\ntheta)] + c\,,
\end{equation}
where $c$ is a normalizing constant and $\E_{\Z_j \mid j \neq i}$ denotes the expectation under all the variables $\Z_j$ with respect to distributions $Q(\Z_j)$, for $j\neq i$. Since by the assumption described in Equation \eqref{eq:Q_dist}, the distribution $Q(\Z_i)$ is in the class of Multinomial distributions and $\Z_1,\dots,\Z_n$ are independent, we have $\mathbb{E}_{\Z_m \mid m\neq i}[Z_{ja}] = \tau_{ja}$, for $j\neq i$.
 Thus,
\begin{equation}
\begin{split}
        \log Q(\Z_i) & =  \mathbb{E}_{\Z_m \mid m \neq i}[\log\mathbb{P}(\A\mid\Z,\ntheta)+\log\mathbb{P}(\Z\mid\npi] + c \\
        & = \mathbb{E}_{\Z_m \mid m \neq i}\left[\sum_{\substack{j\neq i}}\sum_{a,b=1}^{K}Z_{ia}Z_{jb} \log f_{ab}(A_{ij})\right ] + \mathbb{E}_{\Z_m \mid m \neq i}\left[\sum\limits_{a=1}^{K} Z_{ia}\log(\pi_{a})) \right ]+ c\\
        &= \sum_{\substack{j\neq i}}\sum_{a,b=1}^{K}Z_{ia}\tau_{jb} \log f_{ab}(A_{ij}) +\sum\limits_{a=1}^{K} Z_{ia}\log(\pi_{a}) + c\\
         & = \sum_{a=1}^{K}Z_{ia} \left ( \sum_{j\neq i} \sum_{b=1}^{K}\tau_{jb}\log f_{ab}(A_{ij}) + \log(\pi_{a})  \right )+ c\\
    & = \sum_{a=1}^{K}Z_{ia} \log (\tau_{ia})+c\,,
\end{split}
\end{equation}
where the last equality follows from the fact that  $\Z_i \sim \mathcal{M}(1, \nt_i)$. In this way, we have that
\begin{equation}
        \log\tau_{ia}\propto  \sum_{\substack{j\neq i}} \sum_{b=1}^{K}\tau_{jb}\log f_{ab}(A_{ij}) + \log(\pi_{a}).
        \label{est8}
\end{equation}

This leads us to Equation \eqref{fixed_tau}. Since the value of $\tau_{ia}$ depends on $\nt_{j}$, $j \neq i$, it is necessary to implement a fixed-point iteration algorithm. To do so, at step $t$ of the fixed-point algorithm, the values $(\nt_1^{t+1}, \dots, \nt_n^{t+1})$ are computed using $(\nt_1^{t}, \dots, \nt_n^{t})$ and \eqref{fixed_tau}, and then each $\nt_i^{t+1}$ is normalized.

\section{Variational M-step}\label{sec:M-step}

In the M-step of \eqref{eq:VEM} we need to find the estimators of the parameters $\nLambda,\p, \npi, \nmu$ and $\nnu$ which maximize the ELBO. First, using the decomposition of the complete data log-likelihood \eqref{eq:vero} we write

\begin{equation}\label{eq:elbo_tau}
\begin{split}
    \mathfrak{L}(Q,\ntheta) &=  \mathbb{E}_Q[\log \mathbb{P}(\A,\Z\mid \ntheta)] - \E[\log Q(\Z)]\\
   &= \sum_{\substack{i\neq j}}\sum_{a,b=1}^{K}\tau_{ia}\tau_{jb} \log f_{ab}(A_{ij}) + \sum\limits_{i=1}^{n}\sum\limits_{a=1}^{K} \tau_{ia}\log(\pi_{a}) +  \sum_{a=1}^{K} \tau_{ia}\log\tau_{ia}\,.
   \end{split}
\end{equation}

Observe that maximizing the ELBO with respect to parameters $\ntheta$ is equivalent to maximizing the first two terms in \eqref{eq:elbo_tau}. The ELBO function \eqref{eq:elbo_tau} is maximized with respect to $\npi$ considering the constraint that $\sum_{a=1}^{K}\pi_{a}=1$. So, by means of the Lagrange Multipliers we obtain that
\begin{equation}\label{est_pi}
    \hat{\pi}_a=\frac{\sum_{i=1}^{n}\tau_{ia}}{\sum_{a=1}^{K}\sum_{i=1}^{n}\tau_{ia}}=\frac{1}{n}\sum_{i=1}^{n}\tau_{ia}.
\end{equation}
Observe that $\hat{\pi}_a$ only depends on the probabilities given by $\nt_i$, $1\leq i\leq n$.

To maximize \eqref{eq:elbo_tau} with respect to $\nLambda$, $\p$, $\nmu$, and $\nnu$, it is only necessary to maximize its first term, which depends on the log-likelihood of the ZIP distribution, $\log f_{ab}(A_{ij})$. Since the maximum of this log-likelihood with respect to the parameters does not have a closed form, it can be computed using numerical methods such as Newton-Raphson. Alternatively, we propose an EM algorithm, which has shown consistency in the simulations presented in Section \ref{sec:simulation}.

Given a network $\A$ originated from the model \eqref{def:sbm_zip} the observed zero entries of $\A$ can be interpreted as the non-existence of an edge or as the existence of an edge of weight zero generated by a Poisson distribution. Thus, we can introduce the binary latent random variables $W^{ab}_{ij}$, $1 \leq i , j \leq n$, $i\neq j$, that indicate which component $A_{ij}$ comes from when $i$ is in community $a$ and $j$ is in community $b$. In this case, $W^{ab}_{ij}=1$ when $A_{ij}$ comes from the Bernoulli component and $W^{ab}_{ij}=0$, when $A_{ij}$ comes from the Poisson component. Thus, we can rewrite \eqref{def:zip} using the latent variables $W_{ij}^{ab}$ as
\begin{equation}\label{eq:complete}
 f_{ab}(a_{ij}) = (p_{ab} \mathds{1}_{\{a_{ij}=0\}})^{W^{ab}_{ij}}\left((1-p_{ab})\frac{(\mu_i\nu_j\lambda_{ab})^{a_{ij}}e^{-\mu_i\nu_j\lambda_{ab}}}{a_{ij}!}\right)^{1-W^{ab}_{ij}}
\end{equation}
$a_{ij}\in \{0,1,\dots\}$.

Let $\W$ be the collection of latent variables $W^{ab}_{ij}$, $1\leq i, j \leq n$, $i\neq j$ and $1 \leq a, b \leq K$. Using the logarithm of the complete likelihood \eqref{eq:complete} in the first term of  \eqref{eq:elbo_tau}, the EM algorithm at step $s+1$ consists of two stages:
\begin{itemize}
    \item \textbf{E-step:} Calculate the expectation with respect to the conditional distribution of the latent variable $\W$ given the network $\A$ and the current parameters estimates at step $s+1$:
     \begin{equation}
     F(\nLambda^{(s)},\p^{(s)},\nmu^{(s)},\nnu^{(s)})= \E_{\W\mid \A,\nLambda^{(s)},\p^{(s)},\nmu^{(s)},\nnu^{(s)}} \left[\sum_{\substack{i\neq j}}\sum_{a,b=1}^{K}\tau_{ia}\tau_{jb} \log f_{ab}(A_{ij})\right].
     \end{equation}
    \item \textbf{M-step:} Find the parameters that maximize the expected value computed in the E-step:
     \begin{equation*}
(\nLambda^{(s+1)},\p^{(s+1)},\nmu^{(s+1)},\nnu^{(s+1)})=\arg\max\limits_{\nLambda,\p,\nmu,\nnu}F(\nLambda,\p,\nmu,\nnu).
     \end{equation*}
\end{itemize}

Taking the expectation of the logarithm of the complete likelihood \eqref{eq:complete} with respect to the conditional distribution of the latent variable $\W$, given the network $\A$ and the parameters, we have that
\begin{equation} \label{eq:esp_elbo}
\small
\begin{split}
       F(\nLambda,\p,\nmu,\nnu)=\sum_{\substack{i\neq j}}\sum_{a,b=1}^{K}&\tau_{ia}\tau_{jb} \left [\alpha_{ij}^{ab,(s)} \log \left ( \frac{p_{ab}}{1-p_{ab}}\right ) \right.\\
      &\quad+\left. \left ( 1- \alpha_{ij}^{ab,(s)} \right )\log \left (\frac{e^{-\mu_i\nu_j\lambda_{ab}}(\mu_i\nu_j\lambda_{ab})^{a_{ij}}}{a_{ij}!}  \right ) +  \log(1-p_{ab}) \right ],
      \end{split}
\end{equation}
where
\begin{equation}\label{eq:passo_em_alpha}
      \begin{split}
          \alpha_{ij}^{ab,(s)}&={\E}_{\W\mid \A,\nLambda^{(s-1)},\p^{(s-1)},\nmu^{(s-1)},\nnu^{(s-1)}}(W^{ab}_{ij})\\
          &= \frac{p^{(s-1)}_{ab}\mathds{1}_{\{a_{ij}=0\}}}{p_{ab}^{(s-1)}\mathds{1}_{\{a_{ij}=0\}}+ (1-p^{(s-1)}_{ab})e^{-\mu_i^{(s-1)}\nu_j^{(s-1)}\lambda^{(s-1)}_{ab}}}.
      \end{split}
\end{equation}

At the M-step, we maximize the function  \eqref{eq:esp_elbo}  with respect to $\p$, $\nLambda$, $\nmu$ and $\nnu$. To maximizes over $\nLambda$, $\nmu$ and $\nnu$ we use the Expectation Conditional Maximization (ECM) algorithm \citep{Meng_ECM1993}. In this method, we replace the maximization of $F(\boldsymbol{\Lambda}, \boldsymbol{p}, \boldsymbol{\mu}, \boldsymbol{\nu})$ with a series of conditional maximizations. We first update the parameters $\boldsymbol{p}$, followed by the update of $\boldsymbol{\Lambda}$ given $\boldsymbol{\mu}$ and $\boldsymbol{\nu}$. Then, we update $\boldsymbol{\mu}$ given $\boldsymbol{\Lambda}$ and $\boldsymbol{\nu}$, and finally update $\boldsymbol{\nu}$ given $\boldsymbol{\Lambda}$ and $\boldsymbol{\mu}$.  The updates of the parameters $\p, \nLambda, \nmu$ and $\nnu$ are given by Equations \eqref{eq:p_hat},\eqref{eq:lambda_hat}, \eqref{eq:mu_hat} and \eqref{eq:nu_hat}, respectively. 

Observe that the expected out-strength of node $i$ conditioned on $\Z$ and on the latent class $\W$ is given by
\begin{equation}
    \E_{Q,\W}\left(\E\left(\sum\limits_{j=1}^nA_{ij}\mid \Z,\W\right)\right) = \mu_i\sum_{j=1}^n\sum_{a,b=1}^K\tau_{ia}\tau_{jb}\nu_j\lambda_{ab}\left ( 1-\alpha_{ij}^{ab} \right )
\end{equation}
and the in-strength of node $j$ is given by
\begin{equation}
    \E_{Q,\W}\left(\E\left(\sum\limits_{i=1}^nA_{ij}\mid \Z,\W\right)\right) = \nu_j\sum_{i=1}^n\sum_{a,b=1}^K\tau_{ia}\tau_{jb}\mu_i\lambda_{ab}\left ( 1-\alpha_{ij}^{ab} \right )\,.
\end{equation}
Thus, the update of the parameters $\hat\mu_i$ and  $\hat\nu_i$  given in \eqref{eq:mu_hat} and \eqref{eq:nu_hat}, respectively, are equivalent to replace the expected out-strength and in-strength by their observed values.

\section{ICL}\label{sec:icl}

The ICL is a model selection criterion that has been proposed by \cite{Biernacki2000} to select the number of mixture components in a finite mixture model. \cite{daudin2008mixture} adapted the ICL criterion to estimate the number of communities in the SBM, by approximating the complete data likelihood through integrating out the parameters, that is,

\begin{equation}
  \text{ICL}_k(\A) = \P(\A,\Z\mid k) = \int\limits_{\Theta^k} \P(\A,\Z \mid \ntheta,k)\nu(\ntheta\mid k)d\ntheta
\end{equation}
where $\nu(\cdot\mid k)$ is a prior distribution on the parameters $\ntheta=(\npi,\p,\nLambda,\nnu,\nmu)$ and $\Theta^k$ is the parametric space. Assuming that the prior distribution factorizes and using Lemma 3.1 of \cite{Biernacki2000} we have that 
\begin{equation}\label{eq:log_icl}
  \text{ICL}_k(\A) = \log\P(\A\mid \Z, k) + \log\P(\Z\mid k)\,.
\end{equation}

The first term in the right-hand side of \eqref{eq:log_icl} can be calculated using a BIC approximation. Since in $\A$ we have $n(n-1)$ random variables and the number of parameters in $\p$, $\nLambda$, $\nmu$ and $\nnu$ is $k(k+1)/2$, $k(k+1)/2$, $n$ and $n$, respectively, we have 
\begin{equation}\label{eq:cond_log}
\log\P(\A\mid \Z, k) = \max\limits_{\p,\nLambda,\nmu,\nnu}\log\P(\A\mid \Z,\p,\nLambda,\nmu,\nnu, k) - \left( \frac{k(k+1)}{2} + n\right)\log(n(n-1)).
\end{equation}

Assuming a Dirichlet($1/2,\dots,1/2$) prior for $\npi$ and we get
\begin{equation}\label{eq:marginal_Z}
\begin{split}
 \log\P(\Z\mid k) &= \log\Gamma(k/2)-k\log \Gamma(1/2) + \sum\limits_{a=1}^k\log \Gamma(n_a+1/2)-\log \Gamma(n+k/2)\,.
\end{split}
\end{equation}
where $n_a=\sum_{i}Z_{ia}$ is the number of nodes in community $a$. Since $\Z$ is unknown, we replace it with the estimated community membership $\hat\Z$ and its corresponding estimated counts $\tilde n_a = \sum_{i}\hat Z_{ia}$ in \eqref{eq:marginal_Z}. Assuming that $\tilde n_a$ are sufficiently large, we can apply the Stirling approximation to the Gamma function, as used in Proposition 8 of \cite{daudin2008mixture}, resulting in
\begin{equation}
\begin{split}
\log\P(\Z\mid k)  = \max\limits_{\npi}\P(\hat\Z\mid \npi,k)-\frac{k-1}{2}\log n.
\end{split}
\end{equation}

Thus, the ICL is
\begin{equation}
\begin{split}
\text{ICL}_k(\A)& = \max\limits_{\ntheta}\log\P(\A,\hat\Z\mid \ntheta) - \left(\frac{k(k+1)}{2}+n\right)\log(n(n-1))-\frac{k-1}{2}\log n.
\end{split}
\end{equation}

\bibliographystyle{apa}

\bibliography{references}

\end{document}